\documentclass[twocolumn,showpacs,preprintnumbers,amsmath,amssymb]{revtex4}

\usepackage{graphicx}
\usepackage{dcolumn}
\usepackage{bm}

\begin{document}

\title{Field-induced transition from parallel to perpendicular parametric pumping\\ for a microstrip transducer} 

\author{T. Neumann}
 \email{neumannt@physik.uni-kl.de}
\author{A. A. Serga}
\author{V. I. Vasyuchka}
\author{B. Hillebrands}
\affiliation{%
Fachbereich Physik and Forschungszentrum OPTIMAS\\
Technische Universit\"at Kaiserslautern, 67663 Kaiserslautern, Germany}%

\date{\today}

\begin{abstract}
Microstrip transducers used for the excitation of spin waves in magnetic films possess two
characteristic properties: high spatial localization of the microwave magnetic field and the presence of
field components parallel and perpendicular to the bias field. Here, the effects of these features on
the process of parametric pumping are presented. By microwave measurements of the spin-wave instability
threshold a transition from parallel pumping to perpendicular pumping at the critical field $H_{\rm c}$
with the minimal threshold is observed. This transition is accompanied by a sharp threshold increase
above the critical field due to the spatial confinement of the pump region.
\end{abstract}

\maketitle

Microstrip and coplanar transducers have become a standard instrument in magnetism. For the forced
excitation of spin waves they posses several advantages. They can be applied to thin films and
nano-scaled magnetic structures. When used to realize parametric pumping, the pump region is accessible
with optical techniques and the applied microwave power is concentrated into a very high pump field.

Parametric pumping by itself plays an important role in experiments on fundamental properties of
magnetic excitations as well as in applications since it allows the amplification, shaping and
processing of microwave signals \cite{Ser07, Ser05, Mel04}.
One of the most exciting applications of this technique was the recent observation of Bose-Einstein
condensation of magnons at room temperature \cite{Dem06}.

The original theory on parametric pumping was developed by Suhl \cite{Suh57} and Schl\"omann
\cite{Sch60} who considered microwave pump fields oriented perpendicular and parallel to the bias
magnetic field, respectively. In the first case, known as subsidiary absorption, the pump field excites
a uniform precession which couples to spin-wave modes. For parallel pumping, the pump field directly
amplifies spin-wave modes. As for subsidiary absorption this is a threshold processes which only sets in
above a certain threshold pump field. The theoretical predictions for the threshold value
were successfully tested 
using cavity resonators which create a spatially uniform pump field \cite{Wie95, Wet83, Liu82}.

The pump field created by microstrip and coplanar transducers, however, is strongly localized and
non-uniform. This affects the pump process not only in the beneficial ways mentioned above. It is, for
instance, well known that the localization plays an important role in the amplification of traveling
spin waves \cite{Mel99, Gor74}. But, so far the general question how the use of a microstrip transducer
affects the pump process, remains unanswered. In particular, the magnetic pump field around the
transducer has a component oriented perpendicular to the bias magnetic field which is usually neglected
in experiments.

\begin{figure}[b]
\includegraphics[height = 38ex]{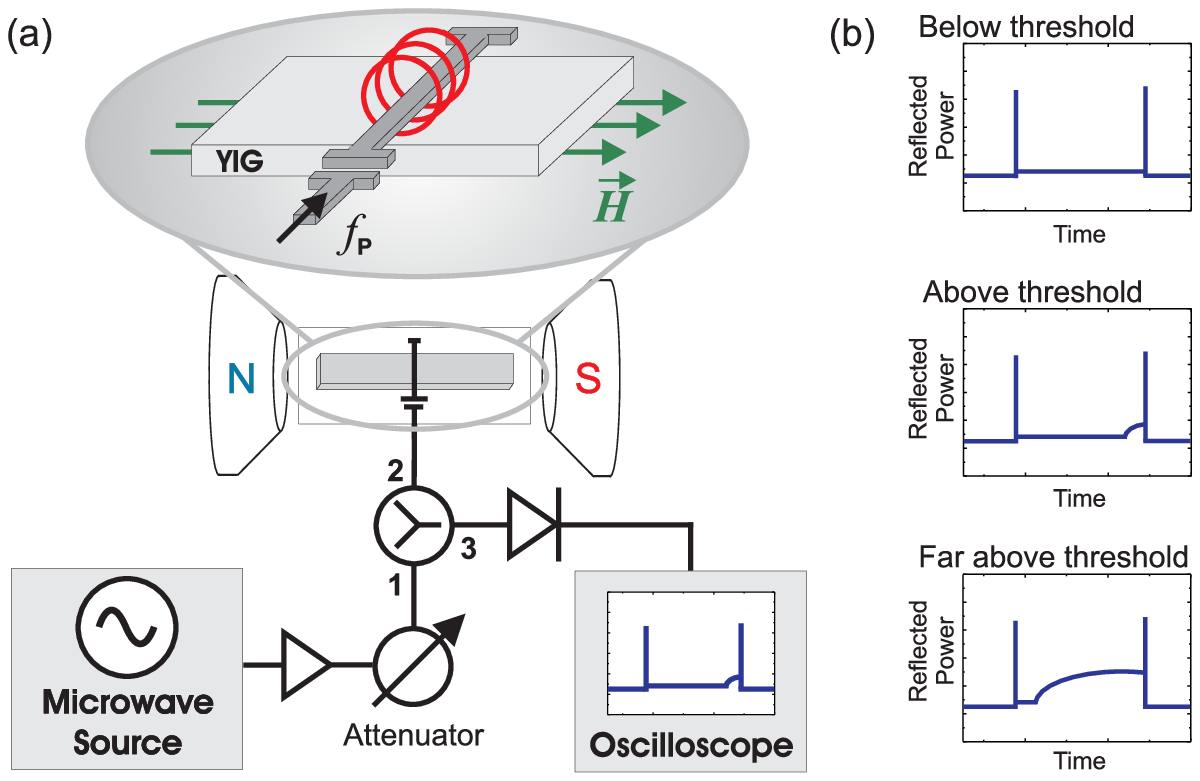}
\caption{\label{fig:aufbau} (Color online) (a) Sketch of the experimental setup. (b) Schematic
oscilloscope pictures for different pump powers}
\end{figure}

To answer the question if this assumption is feasible, the threshold of parametric instability for a
microstrip transducer is investigated. It is shown that below the critical field $H_{\rm c}$ which
corresponds to the minimal threshold power and is given by $H_{\rm c} = - 2\pi M_{\rm S} + \sqrt{(2\pi
M_{\rm S})^2 + (\omega_{\rm p}/(2\gamma))^2}$, where $M_{\rm S}$ is the saturation magnetization,
$\omega_{\rm p}$ is the pump frequency and $\gamma$ is the gyromagnetic ratio, pure parallel pumping is
achieved. In this case, the amplified spin waves propagate along the microstrip transducer and are not
affected by the spatial localization of the pump field. Above the critical field, the spatial
confinement of the pump region leads to an abrupt increase of the threshold because the amplified spin
waves leave the pump region. This increased threshold is accompanied by a transition from parallel to
perpendicular pumping. At bias magnetic fields $H \geq 1.2~H_{\rm c}$, the parametric process at the
threshold is completely determined by the perpendicular component of the magnetic pump field.

We observed the influence of the perpendicular field component already at microwave pump powers around
$10~{\rm mW}$. The efficiency of the amplification by the perpendicular field component is noteworthy if
one takes into account that the frequency of the forced spin oscillations, which serve as a pump source
for the parametrically driven spin waves, coincides with the frequency of the external microwave signal
and lies in our case around 1500 times the resonance linewidth out of resonance. This has to be kept in
mind for numerous experiments with microstrip transducer, where perpendicular pumping could
unintentionally occur.

Fig.~\ref{fig:aufbau}(a) shows the experimental setup. A $20~\mu{\rm s}$ long microwave pulse with a
carrier frequency $\omega_{\rm p}/(2\pi)= 14.14~{\rm GHz}$ was generated every $1~\rm{ms}$. Its power
was adjusted by a power amplifier and an attenuator. The signal was sent to a half-wavelength microstrip
resonator placed on a tangentially magnetized yttrium iron garnet film of $7.8~\mu{\rm m}$ thickness grown in (111)-direction. 
A Y-circulator passed the reflected signal to a detector.

The reflected signal depends significantly on the applied pump power (Fig.~\ref{fig:aufbau}(b)). Below
the threshold, almost no signal is reflected due to the good adjustment of the pump resonator. As soon
as the applied pump power exceeds the threshold power a kink at the end of the reflected signal appears
(middle panel of Fig.~\ref{fig:aufbau}(b)). This is a consequence of the change in the quality factor
and the dynamic detuning of the resonator caused by the excitation of spin waves \cite{Zak74}. In the
experiment, the threshold pump power was determined when the kink appeared after $20~\mu{\rm s}$ at the
end of the pump pulse.


\begin{figure}[t]
\includegraphics[height = 82ex]{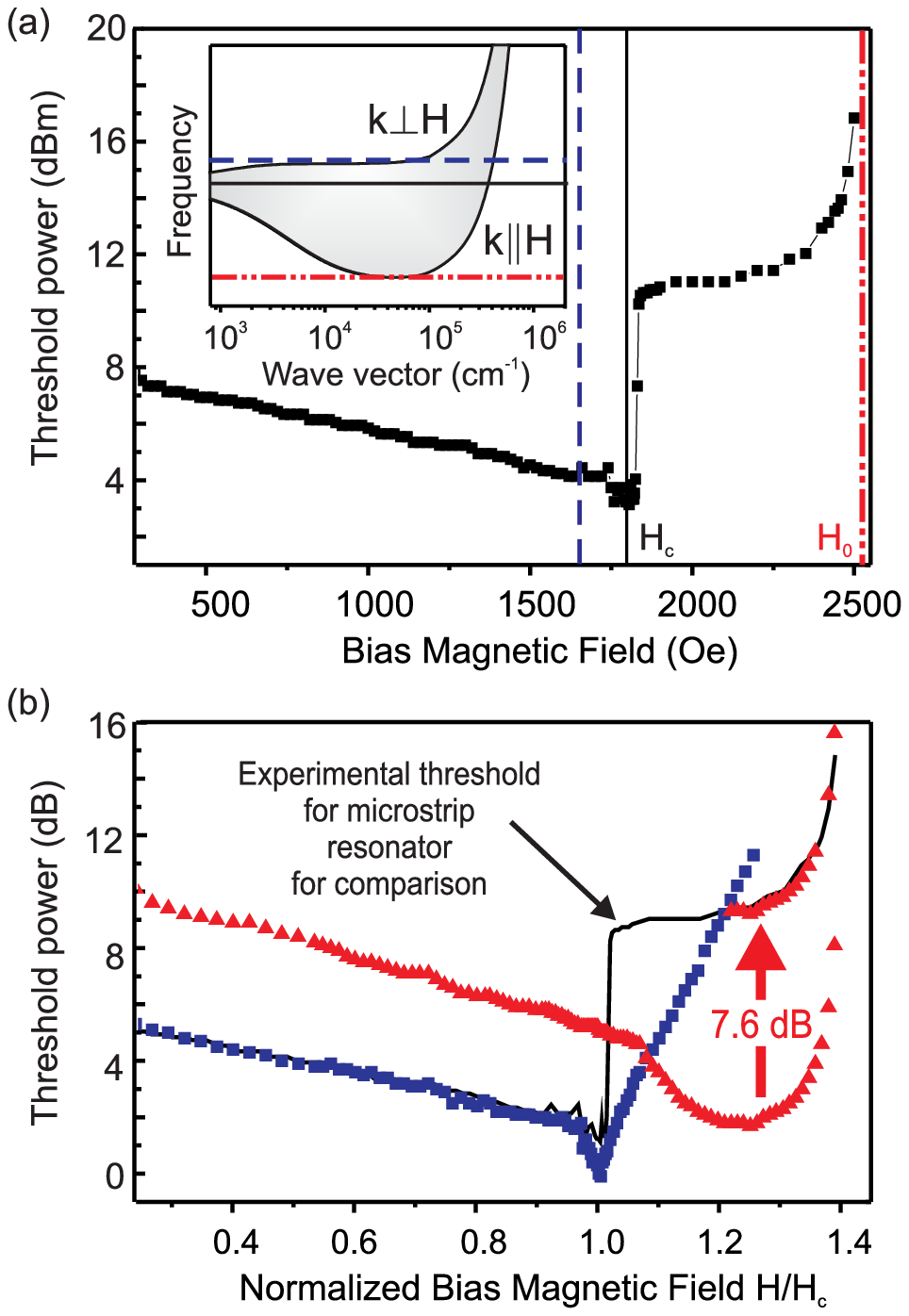}
\caption{\label{fig:results} (Color online) Threshold pump power for (a) microstrip resonator and (b)
dielectric resonator in parallel ($\blacksquare$) and perpendicular ($\blacktriangle$) alignment. The
inset in (a) shows a sketched spin-wave spectrum. The horizontal lines in it correspond to the
respective vertical lines in the main panel and indicate the relative spectral position of the pumped
spin waves.}
\end{figure}

The measured data on the threshold pump power 
is shown in Fig.~\ref{fig:results} (a). In general, the data matches previously reported results with
cavities: (i) When the pump frequency coincides with twice the frequency of ferromagnetic resonance the
threshold power is minimal. The corresponding bias field is defined as $H_{\rm c}$. (ii) For fields
below $H_{\rm c}$ the threshold power decreases with increasing applied bias field. (iii) For fields
above $H_{\rm c}$ the threshold power increases with increasing applied bias field. It grows towards
infinity as the bias field approaches $H_0 = \omega_{\rm p}/(2\gamma)$ and half the pumping frequency
leaves the spin-wave band (see inset in Fig.~{\ref{fig:results}(a)).

The specialty of the data presented here is a sharp jump which occurs just above $H_{\rm c}$. Within a
change of the bias field of $15~{\rm Oe}$ the threshold increases more than 4 times ($7~{\rm dB}$)
before it quickly levels off.

To interpret the observed data consider
\begin{equation} \label{Equ-Threshold}
h_{\rm thr} = \min_{k} \frac{\Gamma_k}{V_k}
\end{equation}
describing the threshold field $h_{\rm thr}$ which depends on the ratio of the spin-wave relaxation
frequency $\Gamma_k$ and the coupling constant $V_k$ of the pump field to the spin wave with wave vector
$k$ \cite{Mel96}.

At the threshold, only one group of spin waves, which satisfies $\omega_k = \omega_{\rm p}/2$ and
simultaneously has the minimal ratio $\Gamma_k / V_k$, is amplified. The decrease of the threshold for
low fields is caused by a decrease in the respective spin-wave relaxation frequency $\Gamma_k$
\cite{Sch60}. 
Small jumps in the threshold are due to the finite film thickness which results in a discrete character
of the spectrum \cite{Kal84, Kos95}.

Below $H_{\rm c}$ the spin waves with the lowest threshold propagate at an angle $\varphi_k =
90^{\circ}$ with respect to the bias field because their coupling $V_k^{\parallel} \sim \sin^2
\varphi_k$ to the parallel pump field is at a maximum. Such spin waves with frequency $\omega_k =
\omega_{\rm p}/2$ do not exist for bias fields above $H_{\rm c}$. Hence, above $H_{\rm c}$ the angle
$\varphi_k$ of the amplified spin waves is smaller than $90^{\circ}$, the coupling $V_k^{\parallel}$ is
weaker and the threshold, consequently, higher. In particular, these spin waves possess a non-zero group
velocity component parallel to the bias field. Therefore, they leave the area of amplification after a
finite time which leads to an increased damping $\Gamma_k$. The smaller the pump region, the more
striking this effect becomes as has been shown in \cite{Mel87} with dielectric resonators of different
sizes. In the
investigated case of a microstrip resonator, this effect is expected to be especially pronounced. 

The observed leveling of the threshold is directly related to the influence of the perpendicular pump
field component. For the perpendicular component the coupling $V_k^{\perp}\sim \sin(2\varphi_k)$ to spin
waves of angle $\varphi_k \approx 45^{\circ}$ is much more efficient than for the parallel field
component. For this reason, the perpendicular pump field component starts to govern the processes at the
threshold above $H_{\rm c}$.

As the applied bias magnetic field approaches $H_0$ both coupling constants $V_k^{\parallel}$ and
$V_k^{\perp}$ decrease leading to a steady increase in the threshold power.

The interpretation was tested by comparing the threshold observed for the microstrip
resonator with the threshold of parametric instability for a dielectric resonator.

For this purpose the YIG-sample was placed inside a $4.2~\times~3.1~\times~1.5~{\rm mm}^3$ dielectric
resonator. Resonator and sample were then fixed inside a waveguide which was on one end shortened with
an adjustable piston and on the other end connected to the setup shown in Fig.~\ref{fig:aufbau}.
Measurements were performed with the same duration of the pump pulse once with the resonator field
pointing parallel and once perpendicular to the bias magnetic field which realizes in good approximation
pure parallel and pure perpendicular pumping.

The measurement results are shown in Fig.~\ref{fig:results}(b).
The obtained curves agree well with previously reported experiments and the established theory (see
references above). Two observations are essential:

(i) The threshold for perpendicular pumping falls below the one for parallel pumping at a field $H \leq
1.1~H_{\rm c}$. This illustrates, that the perpendicular coupling coefficient $V_k^{\perp}$ surpasses
the parallel one $V_k^{\parallel}$ in this field range.

(ii) The slope of the threshold curve for parallel alignment of the dielectric resonator agrees with the
one measured with the microstrip resonator for fields below $H_{\rm c}$. At the same time, the threshold
for the perpendicularly aligned dielectric resonator can be brought to coincidence with the microstrip
resonator threshold for fields above $1.2~H_{\rm c}$.

The relative displacement by $7.6~{\rm dB}$ of the threshold curves for parallel and perpendicular
pumping, which is necessary to match the microstrip resonator threshold curve for low and high bias
fields, is caused by the aforementioned spin-wave mediated energy flux from the pump region.

To qualitatively verify the influence of the energy outflow from the pump region on the threshold power,
consider the following estimate: Let $k$ be the wave vector of the excited spin wave group with the
lowest threshold. For a localized, homogeneous pump region of length $l$ in the direction of the
externally applied field, the outflow of spin waves from this region and the relaxation both lead to a
dissipation of the spin-wave energy $E_k$. Hence,
\begin{equation} \label{Equ-Damping}
\frac{d}{dt}E_k = - 2\Gamma_k \cdot E_k - v_k^{\perp} \cdot \frac{dE_k}{dl}  = -2\big( \Gamma_k +
\frac{v_k^{\perp}}{2l}\big) \cdot E_k
\end{equation}
where, $v_k^{\perp}$ denotes the component of the spin-wave group velocity perpendicular to the
microstrip resonator. According to Eq.~(\ref{Equ-Damping}) the limited size of the pump region leads to
an increase in damping by $v_k^{\perp}/2l$ which should be taken into account in
Eq.~(\ref{Equ-Threshold}) for the threshold field. A rigorous deduction \cite{Lvo94} modifies this
result slightly and includes a geometric factor $\pi<b<2\pi$ which depends on the ratio $\Gamma_k l /
v_k^{\perp}$:
\begin{equation} \label{Equ-Lvov}
h_{\rm thr} = \frac{\sqrt{\Gamma_k^2 + (b~v_k^{\perp}/2l)^2}}{V_k}.
\end{equation}
By looking at the ration of threshold fields $h_{\rm thr}^{\rm ms}/ h_{\rm thr}^{\rm dr} \sim
\sqrt{P_{\rm thr}^{\rm ms}/ P_{\rm thr}^{\rm dr}}$ for microstrip and dielectric resonator, the
effective size of the microstrip resonator pump region $l^{\rm ms}$ can be estimated when the bias field
is above $1.2~H_{\rm c}$ and perpendicular pumping takes place.

For $H = 2350~{\rm Oe} = 1.3~H_{\rm c}$, a typical wave number for parametrically excited spin waves is
of the order of $6\times 10^4~{\rm cm}^{-1}$ \cite{Kos95} with a propagation angle $\varphi =
25^{\circ}$. This leads to $v^{\perp} \approx 0.07 {\rm cm}/\mu{\rm s}$. With an assumed spin-wave
damping $\Gamma_k \approx 5\times10^6~{\rm s}^{-1}$ \cite{Mel99} and the ratio of threshold powers from
the experiment $P^{\rm ms}_{\rm thr}/P^{\rm dr}_{\rm thr} = 10^{7.6/10}$ one calculates $l^{\rm
ms}\approx 100~\mu{\rm m}$. Though this is only a rough estimate it is comparable with the width of the
microstrip resonator of $50~\mu{\rm m}$ and seems, therefore, reasonable.

The calculation confirms, that the increased threshold is caused by the energy outflow from the strongly
localized pump region. In general, for higher magnetic field the propagation angle $\varphi$ of the
parametrically excited spin wave group will decrease, however, its wave number $k$ and the group
velocity component $v^{\perp}$ will remain almost unchanged. This explains why the observed upward shift
of the threshold curve is in fact constant over a whole range of fields from $1.2~H_c$ to $1.4~H_c$.

The above interpretation builds on the perpendicular field component. The threshold curve observed for
the parametric generation of spin waves by a microstrip antenna cannot be explained solely by
considering the field component parallel to the bias field. This becomes particularly clear in
Fig.~\ref{fig:results}(b) at fields above $1.2~H_c$. Here, the threshold curve for the parallel aligned
dielectric resonator surpasses the one for the microstrip resonator. This is impossible to explain on
the base of Eq.~(\ref{Equ-Lvov}).

Moreover, the presented explanation for fields above $1.2~H_c$ solely relies on the perpendicular field
component and neglects the presence of the parallel one altogether. Note that, though it exists, it is
localized at a different position in space and should only have a minor influence on the threshold
power, if any. Overall, the threshold for parametric spin-wave generation by a microstrip transducer
can, therefore, be effectively described by using the model of pure parallel pumping for fields below
the critical field $H_c$ and the model of pure perpendicular pumping for fields above $1.2~H_c$.

In conclusion, we have investigated the threshold of parametric spin-wave instability for a microstrip
resonator. The experiment shows a characteristic jump in the threshold curve just above the critical
bias field. The results are consistently explained as an increase in damping due to the outflow of
energy from the strongly localized pump region around the microstrip resonator and the influence of the
perpendicular pump field component. It is possible to change the character of the pumping by a change of
the bias field $H$ from below $H_{\rm c}$, where the microstrip resonator acts as a source of parallel
pumping, to $H$ above $1.2~H_{\rm c}$, where perpendicular pumping (subsidiary absorption) takes place.
The comparatively high efficiency of the spin-wave excitation by
perpendicular pumping even when the pump frequency is far above the frequency of ferromagnetic resonance
should be considered in any experiment using microstrip transducers.

We would like to thank Prof. G.~A. Melkov, Dr. M.~P. Kostylev and Dr. A.~V. Chumak for their helpful
remarks. This work has been financially supported by the Matcor Graduate School of Excellence and the
DFG within the SFB/TRR 49.


%
%
%

\end{document}